\documentclass[iop]{emulateapj}
\usepackage{amssymb}
\usepackage{epsfig}
\usepackage{apjfonts}
\usepackage{amsmath}

\usepackage{color}

\begin{document}

\title{ Star-forming galaxies as the origin of the IceCube PeV neutrinos}
\author{Xiao-Chuan Chang \altaffilmark{1,2}, Ruo-Yu Liu\altaffilmark{1,3},
Xiang-Yu Wang\altaffilmark{1,2}}

\begin{abstract}
Star-forming galaxies, due to their high star-formation rates and
hence large number of supernova remnants {therein}, are huge
reservoirs of cosmic rays (CRs). These CRs collide with { gases}
in the {galaxies} and produce high-energy neutrinos through $pp$
collisions. In this paper, we calculate the  neutrino production
efficiency in star-forming galaxies by considering realistic
galaxy properties, such as the gas density and galactic wind in
star-forming galaxies. To calculate the accumulated neutrino flux,
we use the infrared luminosity function of star-forming galaxies
obtained by {\em Herschel} PEP/HerMES survey recently. The
intensity of CRs producing PeV neutrinos in star-forming galaxies
is normalized with the observed CR flux at EeV (
{1\,EeV=$10^{18}\,$eV}), assuming that {supernova} remnants or
hypernova remnants in star-forming galaxies can accelerate protons
to EeV energies. Our calculations show that the accumulated
neutrino emission produced by CRs in star-forming galaxies can
account for the flux and spectrum of the sub-PeV/PeV neutrinos
under reasonable {assumptions} { on} the CR confinement time in
these galaxies.
\end{abstract}

\keywords{neutrinos- cosmic rays}

\affil{$^1$ School of Astronomy and Space Science, Nanjing University, Nanjing, 210093, China;  xywang@nju.edu.cn \\
$^2$ Key laboratory of Modern Astronomy and Astrophysics (Nanjing University), Ministry of Education, Nanjing 210093, China \\
$^3$ Max-Planck-Institut f\"ur Kernphysik, 69117 Heidelberg, Germany}

\section{Introduction}

The IceCube Collaboration recently announced the discovery of
extraterrestrial neutrinos. With 37 events ranging from 60 TeV to
3 PeV within three years of operation, the excess over the
background atmospheric neutrinos and muons reaches 5.7$\sigma$
\citep{2014PhRvL.113j1101A}. {The
non-detection of events beyond 3 PeV suggests that the neutrino flux follows either a hard power law spectrum with a break above 3 PeV, or an unbroken power law spectrum with a softer index of $\Gamma \simeq 2.2-2.3$} %
\citep{2014PhRvL.113j1101A,2014PhRvD..89h3003A,Winter2014}.
{The sky distribution of these events is consistent with isotropy} %
\citep{2014PhRvL.113j1101A}, implying that an extragalactic origin is
dominant, although a  fraction of them could come from Galactic sources %
\citep{2013ApJ...774...74F,2013PhRvD..88h1302R,2014PhRvD..90b3016L,2014PhRvD..90b3010A,2014PhRvD..89j3002N}%
.

The source of the IceCube neutrinos is still controversial. The proposed
astrophysical sources include starburst and star-forming galaxies %
\citep{2006JCAP...05..003L,2014PhRvD..89h3004L,He2013,2013PhRvD..88l1301M,2014JCAP...09..043T,
2014PhRvD..89l7304A,Wang2014}, gamma-ray bursts
\citep{1997PhRvL..78.2292W,2013PhRvL.111l1102M,2013JCAP...06..030C,Liu&Wang2013}%
, and jets and/or cores of active galactic nuclei (AGNs) %
\citep{2008APh....29....1A,Kalashev2013,Stecker2013, Dermer2014},
newborn pulsars \citep{2014PhRvD..90j3005F} and etc. In this
paper, we focus on the scenario of starburst/star-forming
galaxies, where cosmic rays (CRs) therein collide with dense gases
{ in intersteller medium} and produce neutrinos. The gamma-ray
observations of nearby star-forming galaxies by Fermi Large Area
Telescope (LAT),
including M31, LMC, SMC, M82, NGC 253 and NGC 2146 %
\citep{2012ApJ...755..164A,Tang2014}, have proven that $pp$
collisions occur in such star-forming galaxies, so they are
guaranteed { factories} of high-energy neutrinos.

In the pioneering work of the starburst galaxy scenario, %
\citet{2006JCAP...05..003L} {assume} that CRs in the starburst
{galaxies} lose almost all the energy into pions and calibrate
{the GeV neutrino emissivity} with the synchrotron radio
emissivity. A simple power-law extrapolation is then used to
estimate the neutrino flux at PeV energies. On the other hand, we
\citep{2014PhRvD..89h3004L} studied the PeV neutrino emissions
from star-forming/starburst galaxies, assuming that the intensity
of CRs producing PeV neutrinos in star-forming/starburst galaxies
matches the observed CR flux at EeV. This is motivated by the
possibility that PeV neutrinos could originate from the same
sources responsible for extragalactic CRs
\citep{2014PhRvD..89h3004L}, as PeV neutrinos require $\sim50$ PeV
CRs, which is only one order of magnitude lower than the energy of
the "second knee" of CR spectrum ($4-8\times10^{17}\mathrm{eV}$),
where the transition from Galactic CRs to extragalactic CRs may
occur\footnote{{ In another model, the transition} occurs at the
"ankle" ($\la10^{19}{\rm eV}$) \citep{Katz&Waxman09}, where the
spectral index flattens from -3.3 to -2.7 . }
\citep{Berezinsky2006}. It has been suggested that the remnants of
hypernovae or other peculiar types of supernovae in star-forming
galaxies may accelerate protons
to EeV energies due to their faster ejecta and larger explosion energy %
\citep{2007PhRvD..76h3009W,Budnik2008,Chakraborti2011,Liu&Wang2012}. In %
\citet{2014PhRvD..89h3004L}, we {assume} that all star-forming
galaxies {as well as all starburst galaxies} have uniform
properties {such as gas density}, which {lead} to the same
neutrino production efficiency among all star-forming galaxies,
and among all starburst galaxies. However, the CR intensity and
gas density should be different in galaxies of different
luminosities or types. In this paper, we attempt to improve our
earlier calculation by considering realistic galaxy properties and
the galaxy luminosity function. The luminosity function describes
the relative number of galaxies of different luminosities, as well
as the evolution of a galaxy population with redshift. The
Herschel PEP/HerMES survey has recently provided an estimate of
the IR luminosity function up to $z\sim 4$
\citep{2013MNRAS.432...23G}. It also {enables people} to estimate
the luminosity functions of specific galaxy populations: normal
spiral galaxies, starbursts and star-forming galaxies containing
obscured or low-luminosity AGNs, all of which contribute to the
star-formation rate in the Universe \citep{2013MNRAS.432...23G}.
The gas density in a galaxy is expected to relate to its
star-formation rate and hence to the IR luminosity of the galaxy.
Then, with these galaxy properties known, we are able to calculate
the neutrino fluxes produced in star-forming galaxies of different
luminosities and populations.

In \S 2, we first outline the neutrino production process in
star-forming galaxies. In \S 3, we describe the galaxy parameters
that are needed for calculating the accumulated neutrino flux from
star-forming galaxies. In \S 4, we invoke the luminosity function
to calculate the accumulated neutrino flux produced by all the
star-forming galaxies in the Universe. Finally, we give our
conclusions and discussions in \S 5.

\section{Neutrino production process in star-forming galaxies}

Supernova remnants {(SNRs)}are widely discussed as accelerators of
CRs. Due to high star-formation rates (SFRs) in star-forming
galaxies, {large amount of SNRs} reside in these galaxies and
hence these galaxies are huge reservoirs of CRs. The total energy
of CRs injected into a galaxy per unit time is proportional to the
total SFR of the galaxy, i.e., $L_{p}\propto \mathrm{SFR}$, where
$L_p$ represents the luminosity in CR protons. The total infrared
luminosity of a galaxy is a good tracer for its SFR, and there
exists a widely used relation between the total infrared
luminosity $L_{\text{TIR}}$ and SFR \citep{1998ApJ...498..541K}.
Thus we expect that $L_p\propto L_{\mathrm{TIR}}$, i.e.
\begin{equation}
L_p=C\frac{L_{\mathrm{TIR}}}{L_\odot}\left(\frac{E_p}{1\mathrm{GeV}}\right)^{-p},
\end{equation}
where $C$ is the normalization factor, $L_\odot$ is the bolometric
luminosity of the Sun and $E_p$ is the proton energy in unit of GeV.
$p$ is the index of the proton spectrum ($dn/dE_p\propto E_p^{-p}$), and we
assume $p=2$, as expected from the first-order Fermi acceleration in
blastwaves of SNRs.

Once the accelerated CRs are injected into interstellar medium (ISM),
hadronuclear collisions between CRs and nuclei in ISM would produce charged
pions, which will decay to neutrinos ($\pi ^{+}\rightarrow \nu _{\mu }\bar{\nu}%
_{\mu }\nu _{e}e^{+},\,\pi ^{-}\rightarrow \bar{\nu}_{\mu }\nu _{\mu }\bar{%
\nu}_{e}e^{-}$). On the other hand, CRs can escape a galaxy through
diffusion or galactic wind advection. These two competing processes regulate
the efficiency of the pion-production of CRs which can be described by $%
f_{\pi }=1-\exp \left( -t_{\text{esc}}/t_{\text{loss}}\right) $, where $t_{%
\text{esc}}$ is the escape time of CRs and $t_{\text{loss}}$ is the
energy-loss time of CRs via proton--proton ($pp$) collisions.

The energy-loss time $t_{\text{loss}}$ can be expressed as $\left(
0.5n\sigma _{pp}c\right) ^{-1}$, where 0.5 is the inelasticity factor, $n$
is the particle number density and $\sigma _{pp}$ is the inelastic $pp$
collision cross section. We convert the particle number density to gas
surface density by $\Sigma _{g}=m_{p}nH$, where $m_{p}$ is the mass of proton
and $H$ is the height of the galaxy, the energy-loss time is
\begin{equation}
t_{\text{loss}}=1.4\times 10^{5}\frac{H}{1\text{kpc}}\left( \frac{\sigma
_{pp}}{70\text{mb}}\right) ^{-1}\left( \frac{\Sigma _{g}}{1\text{g cm}^{-2}}%
\right) ^{-1}\text{yr}.
\end{equation}

There are basically two ways for CRs to {escape from a galaxy},
i.e. diffusion and
advection. In the diffusion escape case, {CRs are scattered by small-scale inhomogeneous magnetic fields randomly} and diffuse out of the host galaxy. The diffusive escape time is $t_{\text{diff}}=H^{2}/4D$. Here $D=D_{0}\left( E/E_{0}\right) ^{\delta }$ is the diffusion coefficient, where $D_{0}$ and $%
E_{0}={\rm 3 GeV}$ are normalization factors, and $\delta =0-1$
depending on the spectrum of interstellar magnetic turbulence. The
diffusion time is
\begin{equation}
t_{\text{diff}}=7.5\times 10^{6}\left( \frac{H}{1\text{kpc}}\right)
^{2}\left( \frac{D_{0}}{10^{28}\text{cm}^{2}\text{s}^{-1}}\right)
^{-1}\left( \frac{E_{p}}{3\text{GeV}}\right) ^{-\delta }\text{yr}.
\end{equation}%
Since galaxies with higher IR luminosities are observed to have stronger
magnetic fields \citep{2006ApJ...645..186T}, and the diffusion coefficient is
expected to scale with the CR Larmor radius, these high luminosity galaxies
could have a smaller diffusion coefficient. Thus, we allow lower values of
diffusion coefficient for galaxies with IR luminosity $L_{ \text{TIR}%
}>10^{11}L_{\odot }$ in the calculation, while the diffusion coefficient for
galaxies with IR luminosity $L_{\text{TIR}}<10^{11}L_{\odot }$ is fixed to $%
D_{0,\text{L}}=10^{28}$cm$^{-2}$s$^{-1}$, { the same value as that
of} our Galaxy. The energy dependence of the diffusion coefficient
is also unknown. We assume two cases, one is the commonly-used
value $\delta =0.5$, based on the {measurements} of the CR
confinement time in our Galaxy
\citep{1990A&A...233...96E,2003ApJ...599..582W}, {which is also
consistent with the Kraichnan-type turbulence}. Another choice is
$\delta =1/3$, assuming the {Kolmogorov-type} turbulence.

In the advection escape case, CRs are confined in the galactic wind
and transported outward with the wind in a characteristic timescale
\begin{equation}
t_{\text{adv}}=H/v_{w}=1.8\times 10^{6}\frac{H}{1\text{kpc}}\left( \frac{%
v_{w}}{500\text{km s}^{-1}}\right) ^{-1}\text{yr}
\end{equation}%
where $v_{w}$ is the speed of galactic wind. The real escape time should
involve both effects, and we parameterize it as $t_{\text{esc}}^{-1}=t_{%
\text{diff}}^{-1}+t_{\text{adv}}^{-1}$.

The flux of neutrinos produced in one galaxy is then calculated by the
following analytical formula
\begin{equation}
L _{\nu }\left( E_{\nu }\right) \simeq \int_{E_{\nu }}^{\infty }f_{\pi
}\left( E_{p}\right) L_{p}\left( E_{p}\right) F_{\nu }\left( \frac{E_{\nu }}{%
E_{p}},E_{p}\right) \frac{dE_{p}}{E_{p}},
\end{equation}
where $F_{\nu }\left( E_{\nu }/E_{p},E_{p}\right) $ is the spectrum of the
secondary neutrino emissions given in \citet{2006PhRvD..74c4018K}.

CRs that escape their host galaxies contribute to the
extragalactic CRs observed by us. At higher energies, diffusion
escape timescale is shorter and hence leads to a higher escape
{efficiency} of CRs, which can be estimated as
$f_{\text{esc}}=1-f_{\pi }$. {The spectrum of these escaped CRs
can then be simply expressed as} $f_{\text{esc}}L_{p}$.

\section{Galaxy parameters}

We have seen that the pion-production efficiency depends on
galaxy parameters, such as gas surface density $\Sigma_g$,
galactic wind velocity $v_w$, scale height of the galaxy $H$ and
etc. In this section, we try to give a description of these
parameters, which are needed in calculating the accumulated
neutrino flux in \S 4. Since the infrared luminosity function is
used to characterize the population of galaxies, we try to build
{relations} between these parameters and the total infrared
luminosity of the galaxy.

To determine the gas surface density $\Sigma_g$, we employ the widely used
Kennicutt-Schmidt law, which relates the SFR surface density $\Sigma_{\text{%
SFR}}$ {with} gas surface density, i.e., $\Sigma
_{\text{SFR}}\propto \Sigma _{g}^{1.4}$
\citep{1998ApJ...498..541K}. The Kennicutt-Schmidt law, although
discovered for galaxies in the local Universe, {is proven to be
also valid at high redshift} \citep{2010MNRAS.407.2091G}. In this
paper we use the classic form given by
\citet{1998ApJ...498..541K}, i.e.
\begin{equation}
\Sigma _{\text{SFR}}=(2.5\pm0.7)\times 10^{-4}\left( \frac{\Sigma _{g}}{1M_{\odot }%
\text{ pc}^{-2}}\right) ^{1.4\pm0.15}M_{\odot }\text{
yr}^{-1}\text{ kpc}^{-2}.
\end{equation}%
If the radius of each galaxy is known, we could derive the SFR surface
density $\Sigma _{\text{SFR}}=$SFR/$\pi R^{2}$. Assuming the Chabrier
initial mass function (IMF) \citep{2003PASP..115..763C}, one has
\begin{equation}
\text{SFR}=\frac{L_{\text{TIR}}}{10^{10}L_{\odot }}M_{\odot }\text{ yr}^{-1}.
\end{equation}
Substituting this relation into Kennicutt-Schmidt Law, we get
\begin{equation}
\Sigma _{g}=(7.07\pm1.63)\times 10^{-5}\left(
\frac{\mathrm{L_{TIR}}}{L_{\odot }}\right)
^{\frac{1}{1.4\pm0.15}}\left( \frac{R}{\text{pc}}\right)
^{-\frac{2}{1.4\pm0.15}}\text{g cm}^{-2}.
\end{equation}

There is a correlation between SFR and the total stellar mass for the
majority of star-forming galaxies, which are known as the Main Sequence (MS)
galaxies. The relation is quite tight in the local Universe %
\citep{2010ApJ...721..193P,2012ApJ...757....4P} and also works well at higher
redshift \citep{2007A&A...468...33E,2007ApJ...670..156D,2010A&A...518L..25R}%
. In this paper, we use the SFR-stellar mass relation provided by %
\citet{2010ApJ...718.1001B} and \citet{2010MNRAS.407.2091G}, {i.e., $\text{SFR%
}\left( \text{M}_{\odot }\text{yr}^{-1}\right) =150\left( M_{\star }/10^{11}%
\text{M}_{\odot }\right) ^{0.8}[\left( 1+z\right) /3.2]^{2.7}$ for $z<2.3$ and $\text{SFR%
}\left( \text{M}_{\odot }\text{yr}^{-1}\right) =163\left( M_{\star }/10^{11}%
\text{M}_{\odot }\right) ^{0.8}$ for $z>2.3$ up to $z=4$, since
the redshift evolution flattens above $z\sim 2-2.5$.}

While the MS galaxies {follow the SFR-stellar mass relation
above}, some galaxies have much higher efficiencies in
transforming gas to stars, so they have higher SFRs given the same
stellar mass. These outliers generally follow another linear
relation between SFRs and stellar masses with an offset of a
factor of a few from the MS one. The offset can be measured by,
namely, the specific star formation rate (sSFR), which is defined
as SFR$/M_{\star }$. Galaxies with higher sSFR are thought to be
off-MS galaxies and in a merger mode. According to
\citet{2013MNRAS.432...23G}, normal spiral galaxies, starburst
galaxies and SF-AGNs(spiral) are thought to be mostly on-MS
galaxies, while SF-AGNs(SB) are thought to be off-MS galaxies. In
our calculation, the SFR-stellar mass relation is increased by 0.6
dex for off-MS galaxies. For the Chabrier IMF, the relation
between {the total stellar mass} and the total infrared luminosity
can be summarized as
\begin{equation}
M_{\star }=\left\{
\begin{array}{cc}
10^{11}\left( \frac{L_{\text{TIR}}}{1.5\alpha \times 10^{12}L_{\odot }}%
\right) ^{1.25}\left( \frac{1+z}{3.2}\right) ^{-3.38} & \left( z<2.3\right)
\\
9\times 10^{10}\left( \frac{L_{\text{TIR}}}{1.5\alpha \times 10^{12}L_{\odot
}}\right) ^{1.25} & \left( z>2.3\right),%
\end{array}%
\right.
\end{equation}%
where $\alpha$ is equal to 1 and 4 for on-MS galaxies and off-MS galaxies
respectively.

The relation between stellar mass and galaxy radius has been studied by
different authors %
\citep{2003MNRAS.343..978S,2011MNRAS.410.1660D,2011ApJ...727....5M,2012ApJ...745...85L,2014MNRAS.444..682C}%
. For local late-type galaxies, \citet{2003MNRAS.343..978S} found a relation
based on the Sloan Digital Sky Survey,
\begin{equation}
R_{\text{SDSS}}=0.1\left( \frac{M_{\star }}{M_{\odot }}\right) ^{0.14}\left(
1+\frac{M_{\star }}{3.98\times 10^{10}M_{\odot }}\right) ^{0.25}\text{kpc}.
\end{equation}%
At high redshift galaxies tend to be more compact, while the scaling still
works \citep{2011MNRAS.410.1660D,2011ApJ...727....5M,2012ApJ...745...85L}. %
\citet{2012ApJ...745...85L} found a redshift-dependent relation,
\begin{equation}
\frac{R}{R_{\text{SDSS}}}\approx \left\{
\begin{array}{cc}
1 & \left( z<1\right), \\
2\left( 1+z\right) ^{-1.07} & \left( z>1\right).%
\end{array}
\right.
\end{equation}%
Combining the SFR-stellar mass relation and the radius-stellar mass relation
above, we can now derive the galaxy radius $R$ from its SFR. As the
star-forming galaxies at high {redshift} are more consistent with triaxial ellipsoids with minor/major axis ratio $\sim 0.3$ \citep{2012ApJ...745...85L}%
, we assume $H=0.3R$.

Galactic-scale gaseous outflows or winds in star-forming galaxies are
ubiquitous at all cosmic epochs %
\citep{1990ApJS...74..833H,2001ApJ...554..981P,2003ApJ...588...65S}.
Such outflows are powered by supernova explosions or other
processes. The dependence of galactic wind speed on the galaxy's
SFR has been studied. For ultraluminous infrared galaxies at low
redshifts,  winds from more luminous starbursts have higher speeds
roughly as $v_{w}\propto $ SFR$^{0.35} $
\citep{2005ApJ...621..227M}. Similar relation {is found in}
star-forming galaxies at $z=1.4$, showing $v_{w}\varpropto $
SFR$^{0.3}$ {with an error} in $v_w$ being $34\%$
\citep{2009ApJ...692..187W}. Combining Eq. 1, we get
\begin{equation}
v_{w}\approx 175\left( \frac{\text{SFR}}{M_{\odot }\text{ yr}^{-1}}\right)
^{0.3}\approx 400\left( \frac{L_{\text{TIR}}}{10^{11}L_{\odot }}\right)
^{0.3}\text{km s}^{-1}.
\end{equation}

{Once the galaxy parameters are known, we can calculate the
pion-production efficiency} $f_\pi$ of CRs in galaxies of
different luminosities.Fig.1 shows the efficiency $f_\pi$ for CRs
producing 1 PeV {neutrinos} in a galaxy at $z=1$. One can see that
the $pp$ interaction is {quite inefficient in low IR luminosity
galaxies}, due to low gas densities in these galaxies. As the IR
luminosity increases, the pion-production efficiency increases. It
also shows that the pion-production efficiencies are higher in
off-MS galaxies, {which is due to denser ISM in them}. We also
give the uncertainty of $f_\pi$ in Fig. 1 (the shaded region),
taking into account the uncertainties in the Kennicutt-Schmidt Law
(including the uncertainty in slope) and in the galactic wind
velocity. {We find the uncertainty of $f_\pi$ is about $50\%$
which mainly results from the uncertainty in the slope of the
Kennicutt-Schmidt Law.}

{In our calculation, we assumed  that the column density of gas in
a  galaxy is uniform out to a limiting radius in a galaxy. The
realistic  gas density distribution in a galaxy may have a smooth
gradient outwards. Correspondingly, the CR injection rate, which
traces the SFR, may follow the same distribution. We employ an
exponential density profile found by \citet{2013ApJ...764L..31K}
to re-calculate  the pion-production efficiency and find that the
overall efficiency is decreased by a factor of about $30\%$. }

The pion-production efficiency also depends on the energy of CRs.
As the diffusion escape is faster at higher energy while the
advection and $pp$ interaction timescales are energy-independent,
the pion-production efficiency {would break at some energy and}
then decreases as the energy of CRs increases. As a result, the
escape efficiency for CRs, $f_{\text{esc}}=1-f_{\pi }$, increases
with energy. {That means} CRs above 1 EeV are able to escape
almost freely from their host galaxies and contribute to the
observed flux of extragalactic CRs.

\section{Accumulated neutrino flux with normalization to EeV CRs}

In this section, we compute the accumulated neutrino flux { by}
adopting the Herschel PEP/HerMES luminosity function
\citep{2013MNRAS.432...23G}. Herschel is the first telescope that
allows to detect far-IR population up to $z\simeq 4$.
\citet{2013MNRAS.432...23G} estimate the luminosity functions of
different galaxy populations including normal spiral galaxies,
starbursts and star-forming galaxies containing obscured or
low-luminosity AGNs. The galaxy classification is based on IR
spectra, where those that have far-IR excess with significant
ultraviolet extinction are classified as starbursts and those that
have mid-IR excess are classified as galaxies with obscured or
low-luminosity AGNs (SF-AGN). The SF-AGN family includes Seyferts,
LINERs and ULRIGs containing AGNs.  SF-AGNs are further divided
into two sub-classes: SF-AGN(SB) that resembles starburst galaxies
and SF-AGN(spiral)that resembles normal spiral galaxies. This
family, although containing AGNs, is dominated by star-formation
but not by AGN processes. The accumulated neutrino flux is the sum
of the contribution from all the galaxies throughout the whole
universe, i.e.,
\begin{equation}
\begin{split}
E_{\nu}^{2}\Phi _{\nu _{i}}^{\text{accu}}& =\frac{E_{\nu}^{2}c}{4\pi }\int_{0}^{z_{\max
}}\int_{L_{\text{TIR},\min }}^{L_{\text{TIR},\max }} \\
& \frac{\sum\nolimits_{i}\phi _{i}\left( L_{\text{TIR}},z\right) L_{\nu _{i}}[(1+z)E_{p},L_{\text{TIR}}]}{H_{0}\sqrt{(1+z)^{3}\Omega _{M}+\Omega
_{\Lambda }}}dL_{\text{TIR}}dz.
\end{split}%
\end{equation}%
where $\phi _{i}(L_{\text{TIR}},z)$ is the luminosity function for each
galaxy family $i$, $H_{0}=71\mathrm{\,}$km\thinspace s$^{-1}$\thinspace Mpc$%
^{-1}$, $\Omega _{M}=0.27$, $\Omega _{\Lambda }=0.73$.

The luminosity function of certain {class of galaxies} (denoted by
the subscript $i$) can be generally {described as}
\begin{equation}
\phi _{i}(L_{\text{TIR}},z)=\phi ^{\ast }\left( \frac{L_{\text{TIR}}}{%
L^{\ast }}\right) ^{1-\alpha }\exp [-\frac{1}{2\sigma ^{2}}\log
_{10}^{2}\left( 1+\frac{L_{\text{TIR}}}{L^{\ast }}\right) ]\,
\end{equation}%
where $L^{\ast }$ evolves as $\left( 1+z\right) ^{k_{L,1}}$ at $z<z_{\text{b}%
,L}$, and as $\left( 1+z\right) ^{k_{L,2}}$ at $z>z_{\text{b},L}$, while $%
\phi ^{\ast }$ evolves as $\propto \left( 1+z\right) ^{k_{\rho ,1}}$ at $%
z<z_{\text{b},\rho }$ and as $\left( 1+z\right) ^{k_{\rho ,2}}$ at $z>z_{%
\text{b},\rho }$. For each population of galaxies, the parameters in
luminosity functions, such as $L^{\ast }$, $\phi ^{\ast }$, $\alpha $, $%
\sigma $, $z_{\text{b},L} $, $k_{L,1}$, $k_{L,2}$, $z_{\text{b},\rho }$, $%
k_{\rho ,1}$, $k_{\rho ,1}$ are provided in Table 8 in %
\citet{2013MNRAS.432...23G}. The number ratio between the  two
sub-classes of SF-AGNs (i.e. SF-AGN(spiral) and SF-AGN(SB))evolves
with redshift, as given in Table 9 of \citet{2013MNRAS.432...23G}.

To compute the accumulated neutrino flux from all star-forming
galaxies, we need to determine the normalization for the CR
intensity at EeV energy in each galaxy, i.e. the factor $C$ in
Eq.1. {We assume that the CRs which escape from these galaxies are
responsible for the extragalactic CR flux at EeV}. As EeV CRs do
not suffer significant attenuation during their propagations to
the Earth, the expected CR flux at $E_{p}$ is
\begin{equation}
\begin{split}
E_{p}^{2}\Phi _{p}^{\text{accu}}& =\frac{E_{p}^{2}c}{4\pi }\int_{0}^{z_{\max
}}\int_{L_{\text{TIR},\min }}^{L_{\text{TIR},\max }} \\
& \frac{\sum\nolimits_{i}\phi _{i}\left( L_{\text{TIR}},z\right) f_{\text{esc%
}}L_{p}[(1+z)E_{p},L_{\text{TIR}}]}{H_{0}\sqrt{(1+z)^{3}\Omega _{M}+\Omega
_{\Lambda }}}dL_{\text{TIR}}dz.
\end{split}%
\end{equation}%
The observed flux at 1 EeV is about $E_{p}^{2}\Phi
_{p}|_{E_{p}=1\mathrm{EeV}}\simeq2\times 10^{-7}\text{GeV
cm}^{-2}\text{ s}^{-1}\text{
sr}^{-1}$, according to several CR experiments such as HiRes %
\citep{2009APh....32...53H} , Auger \citep{2012NIMPA.692...83P},
KASCADE-Grande \citep{2014JPhCS.531a2001C} and TA \citep{2015APh....61...93A}%
. Then we get the normalization factor $C\simeq 2\times 10^{22}$eV$^{-1}$s$%
^{-1}$.

The accumulated neutrino flux can then be obtained with Eq.(13).
Since the galaxy parameters are all determined, there are only two
free parameters, i.e. the diffusion coefficient $D_{0}$ and
$\delta $, which are not well-understood. We study whether the
theoretical flux agrees with the observations under reasonable
values of these two parameters. We find that for $\delta =1/3$,
$D_{0,\mathrm{H}}\simeq10^{27.2}$cm$^{2}$s$^{-1}$ (the
diffusion coefficient for galaxies with high IR luminosity $%
L_{\mathrm{TIR}}>10^{11}L_\odot$) leads to a total neutrino flux
that can fit the IceCube data, as shown in Fig.2. Note that the
diffusion coefficient for galaxies with low IR luminosity
$L_{\mathrm{TIR}}<10^{11}L_\odot$ is fixed to $D_{0,
\text{L}}=10^{28}$cm$^{2}$s$^{-1}$. The figure shows contributions
from various types of galaxies. We find that star-forming galaxies
with obscured AGNs and starburst galaxies contribute significant
fractions of the neutrino flux, while the spiral galaxies
contribute the least. As expected, the neutrino spectrum becomes
softer at high energies where the diffusion time
is shorter than the advection time. The slope becomes steeper than $%
\Phi(E_\nu)\propto E_\nu^{-2.1}$ above PeV energy.

For $\delta =0.5$, a smaller values of $D_{0, \text{H}}\simeq10^{26}$cm$^{2}$s$%
^{-1}$ is needed to fit the IceCube data, as shown in Fig.3. With a large $
\delta$, the neutrino spectrum becomes softer than
$\Phi(E_\nu)\propto E_\nu^{-2.3}$ above PeV energy, which could
explain the non-detection of neutrinos above 3 PeV
\citep{2014PhRvD..89h3003A,Winter2014}. The lower value
of $D_{0, \text{H}}$ leads to a diffusion coefficient of $D=5.7\times10^{29}%
\mathrm{cm^2 s^{-1}}$ at $E_p=100 \mathrm{PeV}$. The confinement
of 100 PeV protons requires $E=eB l_{\mathrm{c}}>100
\mathrm{PeV}$, which leads to a coherence length
$l_{\mathrm{c}}>0.1\mathrm{pc}B_{-3}(E_p/100\mathrm{PeV})$. Thus,
the minimum required diffusion coefficient is   $D({\rm 100
PeV})=(1/3)l_{\mathrm{c}} c=3\times10^{27}\mathrm{cm^2 s^{-1}}$
\citep{2014JCAP...09..043T}. The diffusion coefficient obtained
above in explaining the IceCube data  satisfies this condition.

We calculate the diffuse gamma-ray flux {accompanying the neutrino
emission}, following the approach of \citet{2014ApJ...793..131C}.
The results are also shown in Fig. 2
and Fig. 3 for the cases of {$\delta=1/3$, $D_{0, \text{H}%
}\simeq10^{27.2}$cm$^{2}$s$^{-1}$ and $\delta=0.5$, $D_{0, \text{H}
}\simeq10^{26}$cm$^{2}$s$^{-1}$}, respectively. In the
calculation, we considered the synchrotron loss effect of
electron-positron pairs produced by the absorbed gamma rays in the
galaxies. The strength of the magnetic fields in the galaxies are
assumed to be $B=400\mathrm{\mu G}(\Sigma_g/1\mathrm{g
cm^{-2}})^{0.7}$ \citep{2006ApJ...645..186T,2010ApJ...717..196L}.
We find that the accompanying gamma-ray flux is below the diffuse
isotropic gamma-ray background observed by the Fermi/LAT
\citep{2014arXiv1410.3696T}.

We also study the neutrino flux from star-forming galaxies in
different luminosity ranges. Figure 4 presents the result for
$\delta =0.5$ and $D_{0, \text{H}}\simeq10^{26}$cm$^{2}$s$^{-1}$,
the same parameters {as those in} Figure 3. We can see that most
fraction of the neutrino flux is contributed by the galaxies with
total IR luminosity in the range of $10^{11}-10^{13}L_{\odot }$.
This indicates that the accumulated neutrino flux is dominated by
high-luminosity starburst and SF-AGN galaxies.

Figure 5 shows the flux contributed by on-MS galaxies and off-MS ones
separately. According to their locations on the SFR-stellar mass plane,
normal spiral galaxies, starburst galaxies and SF-AGNs(spiral) are thought
to be mostly on-MS galaxies, while SF-AGNs(SB) are thought to be off-MS
galaxies \citep{2013MNRAS.432...23G}. Fig.5 suggests that on-MS galaxies
dominate the contribution to the total neutrino flux over off-MS galaxies.

In above calculations, we have assumed that the CR diffusion coefficient
in galaxies with IR luminosity $L_{\mathrm{TIR}}<10^{11}L_{\odot }$ is fixed to $D_{0,%
\text{L}}=10^{28}$cm$^{-2}$s$^{-1}$, i.e. the same value as that
of our Galaxy. However, there is observational evidence for a
smaller diffusion coefficient in high-redshift star-forming
galaxies \citep{2013ApJ...772L..28B}. Thus, we recalculate the
neutrino flux by taking
$D_{0,\mathrm{L}}=10^{27}\mathrm{cm^{2}s^{-1}}$ for galaxies with
total IR luminosity $L_{\rm TIR}<10^{11}L_{\odot }$, while keeping
other parameters unchanged. The comparison between these two cases
is shown in Fig.6. As is shown, the total neutrino flux changes
only a little. This is mainly because that low IR luminosity
galaxies contribute sub-dominantly to the total neutrino flux due
to lower pion-production efficiencies.

\section{Discussions and Conclusions}

The proposed scenario above is based on the assumption that
CR protons in star-forming galaxies are accelerated to energy
above 1 EeV. Though we suggest that remnants of hypernovae or
other peculiar types of supernova in these galaxies are  possible
accelerators of these CRs, the estimate of the neutrino flux
presented in this paper does not depend on any specific
accelerator sources. If remnants of normal supernovae in
star-forming galaxies are able to accelerate protons to energy
above 100 PeV due to higher magnetic fields in these galaxies, our
calculations also apply. Nevertheless, the fact that normalizing
the CR intensity in star-forming galaxies at EeV with the observed
CR flux results in a neutrino flux comparable to that observed by
IceCube implies that star-forming galaxies are potential origins
of both IceCube PeV neutrinos and extragalactic EeV CRs.

We note that \citet{2014JCAP...09..043T} have discussed the
possibility that star-forming galaxies are the main sources of
IceCube PeV neutrinos by adopting the IR luminosity function of
\citet{2013MNRAS.432...23G}. They first calculated the diffuse
gamma-ray background produced by star-forming galaxies using the
correlation between the gamma-ray intensity and infrared
luminosity reported by Fermi observations. They then obtained the
PeV neutrino flux assuming that all the gamma-rays are produced by
hadronic $pp$ collisions and used power-law extrapolation to PeV
energy. However, the correlation between gamma-ray luminosities
and infrared luminosities {is based on} observations of nearby
galaxies and it is unclear whether this correlation applies to
high-redshift star-forming galaxies. Differently, we do not rely
on this correlation, but calculate the pion production
{efficiencies} (and hence the neutrino flux) in these galaxies
using the available knowledge about high-redshift star-forming
galaxies.

To summarize, {we have calculated} the neutrino flux produced by
CRs in different populations of star-forming galaxies considering
realistic galaxy properties {and the latest} IR luminosity
functions. By normalizing the CR intensity {from the galaxies}
with the observed flux of EeV CRs, {we have found} that the
accumulated neutrino flux from star-forming galaxies can explain
the IceCube observations of sub-PeV/PeV neutrinos.

\acknowledgments We thank Junfeng Wang and Yong Shi for useful discussions.
This work is supported by the 973 program under grant 2014CB845800, the NSFC
under grants 11273016 and 11033002, and the Excellent Youth Foundation of
Jiangsu Province (BK2012011).

\clearpage

\begin{figure}
\epsscale{.9} \plotone{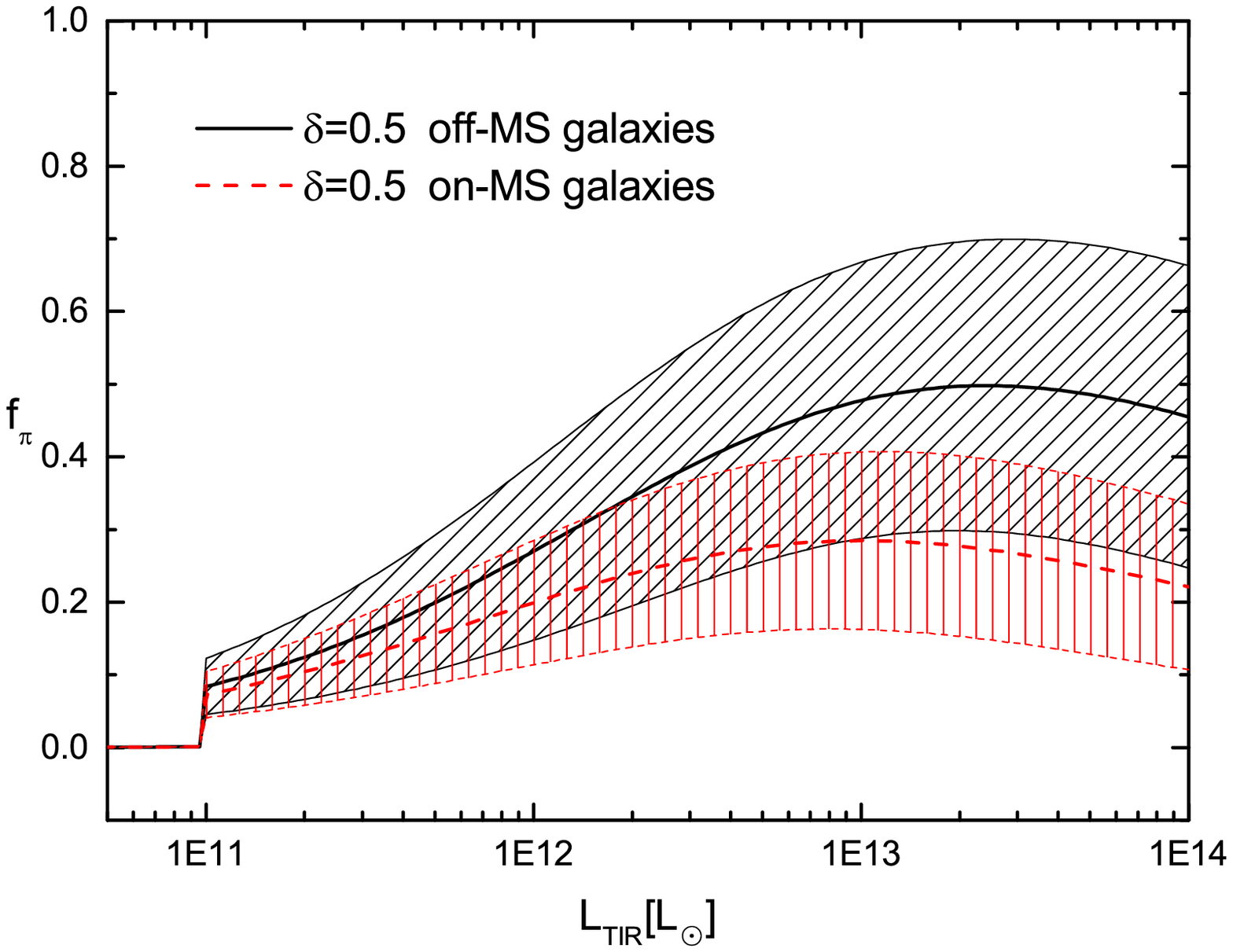} \caption{The pion-production
efficiency $f_\pi$ for CRs producing 1 PeV neutrinos in
star-forming galaxies with different total IR luminosities
(assuming the source is at the redshift $z=1$). The shaded regions
denote the  uncertainties of $f_\pi$ resulted from the
uncertainties in the galaxy properties (see the text for details).
}
\end{figure}

\begin{figure}
\epsscale{.9} \plotone{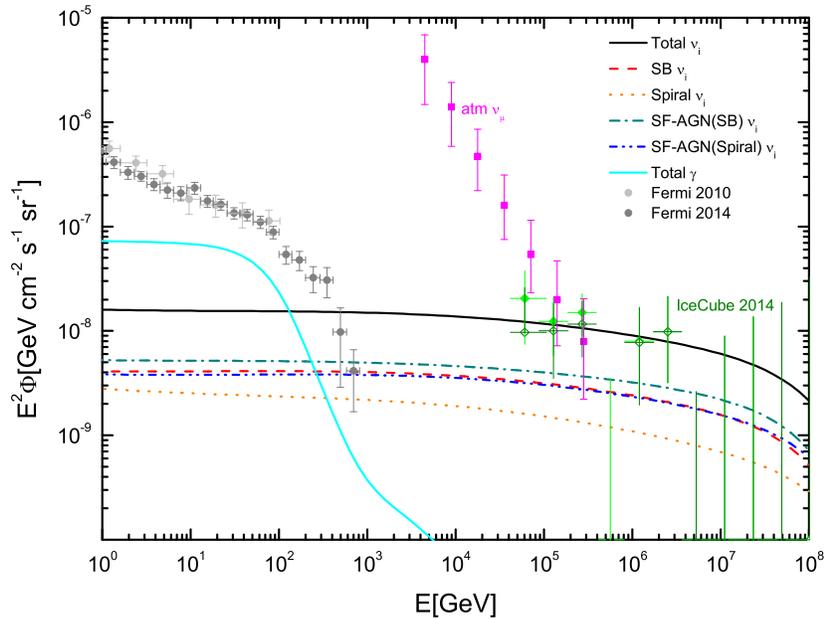}
\caption{Accumulated neutrino flux produced by star-forming galaxies of
different populations. $D_{0,\text{H}}=10^{27.2}$cm$^{2}$s$^{-1}$, $D_{0,\text{L}%
}=10^{28}$cm$^{2}$s$^{-1}$ and $\protect\delta =1/3$ are used in
the calculation. The red, orange, green, blue lines show the flux
contributed by starburst galaxies, normal spiral galaxies, SF-AGNs
(SB) and SF-AGNs (spiral), respectively. And the black line shows
the sum of them. The cyan line shows the diffuse gamma-ray flux
accompanying with the neutrino flux. The extragalactic gamma-ray
background data from Fermi/LAT are depicted as the grey dots. The
atmospheric neutrino data and the IceCube data are also shown.}
\end{figure}

\begin{figure}
\epsscale{.9} \plotone{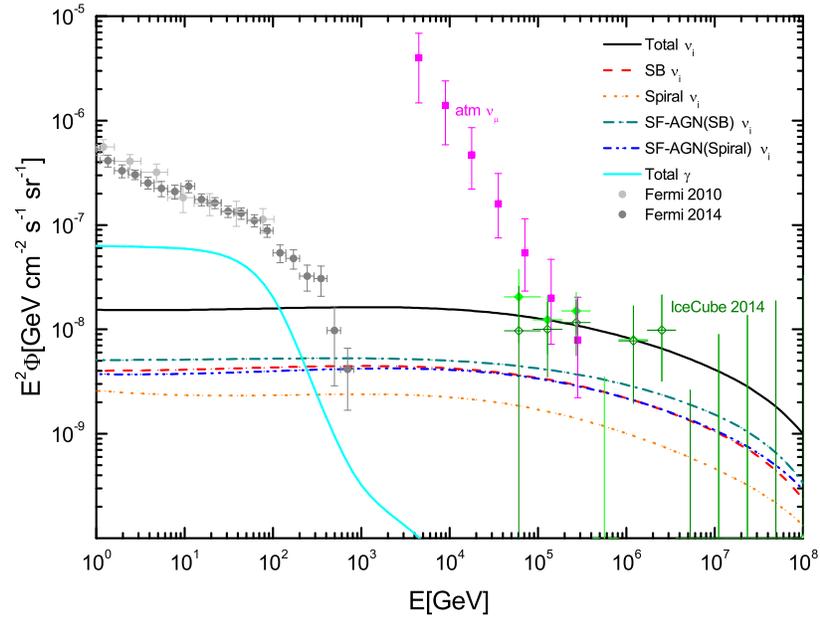} \caption{Same as Figure 2, but with
$\protect\delta =0.5$, $D_{0,\text{H}}=10^{26}$cm$^{2}$s$^{-1}$,
and $ D_{0,\text{L}}=10^{28}$cm$^{2}$s$^{-1}$.}
\end{figure}

\begin{figure}
\epsscale{.9} \plotone{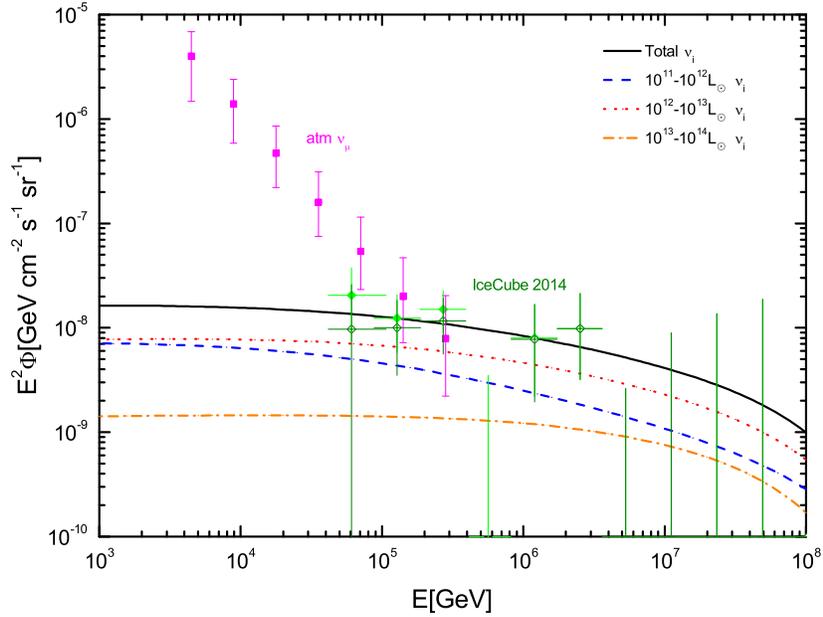}
\caption{Accumulated neutrinos flux from star-forming galaxies of
different IR luminosities.}
\end{figure}

\begin{figure}
\epsscale{.9} \plotone{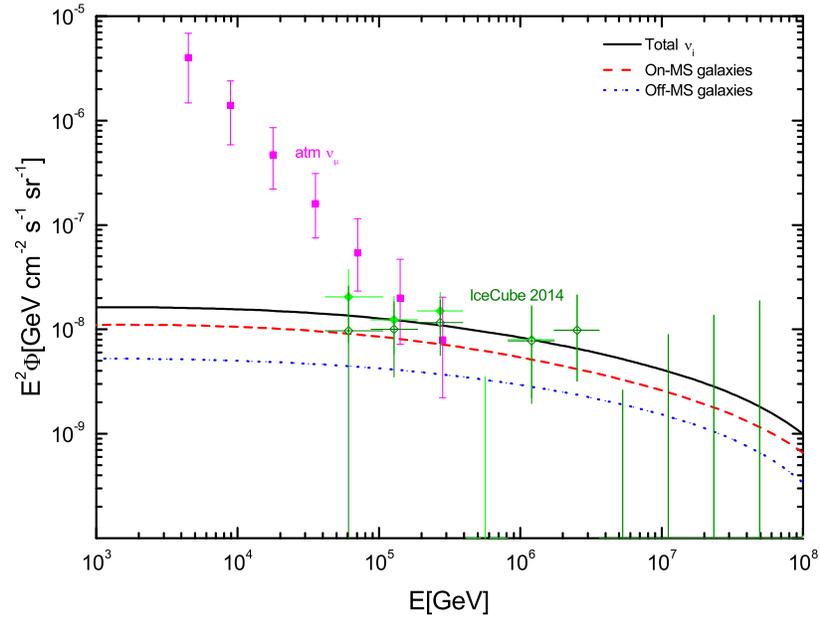} \caption{The same as Figure 3, but
the contribution of the galaxies is divided into two subclasses,
on-MS galaxies and off-MS galaxies. The black line shows the total
flux, while the red and blue lines show the  flux contributed by
off-MS and on-MS galaxies, respectively.}
\end{figure}

\begin{figure}
\epsscale{.9} \plotone{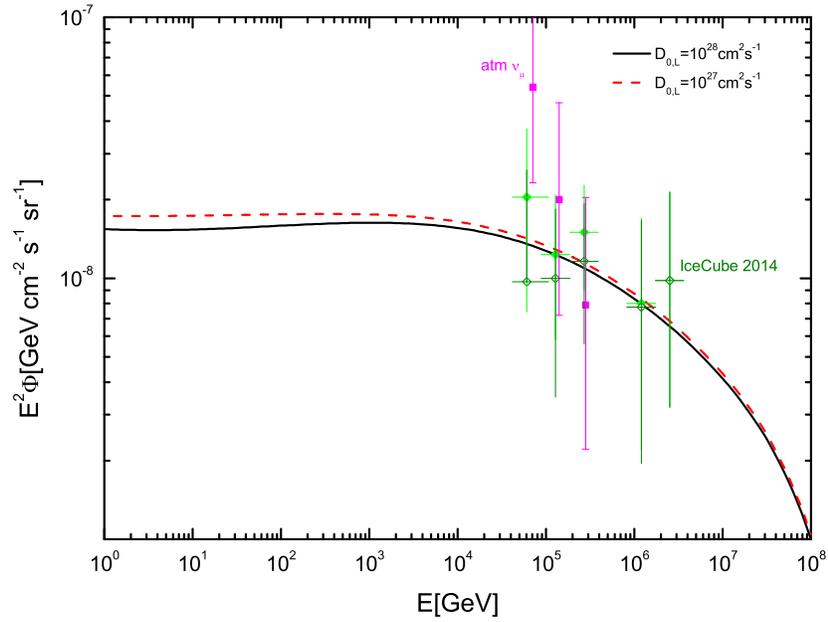} \caption{The accumulated neutrino
flux of star-forming galaxies with different values of $
D_{0,\text{L}}$, while the other parameters remain { unchanged}.
The
black and red lines shows the cases with $D_{0,\text{L}}=10^{28}$cm$^{2}$s$%
^{-1}$ and $D_{0,\text{L}}=10^{27}$cm$^{2}$s$^{-1}$, respectively.}
\end{figure}

\end{document}